\documentclass[a4paper,usenatbib,useAMS]{mnras}

\usepackage{graphicx}   
\usepackage{amsmath}    
\usepackage{amssymb}    

\newcommand{\kms}{\,km\,s$^{-1}$} 
\newcommand{\hmpc}{\,$h^{-1}$\,Mpc\,}
\newcommand{\vparallel}{\mathbin{\!/\mkern-5mu/\!}}

\usepackage[T1]{fontenc}

\usepackage{txfonts}

\title[Coherent motions of cosmic voids]{The sparkling Universe: 
the coherent motions of cosmic voids}

\author[Diego G. Lambas et al.]{\parbox[t]{\textwidth}{\vspace{-1cm}
        Diego G. Lambas\thanks{e-mail: dgl@oac.unc.edu.ar},
				 Marcelo Lares,
				 Laura Ceccarelli,
				 Andr\'es N. Ruiz,
				 Dante J. Paz,
             Victoria E. Maldonado \&
				 Heliana E. Luparello
			         }\vspace{0.2cm}\\
Instituto de Astronom\'{\i}a Te\'orica y Experimental (IATE),
CONICET-UNC\\
Observatorio Astron\'omico, Universidad Nacional de C\'ordoba,
   Argentina.
}

\begin{document}

\date{Accepted XXX. Received XXX; in original form XXX}
\pubyear{2015}
\label{firstpage}    
\pagerange{\pageref{firstpage}--\pageref{lastpage}} \pubyear{XXXX}

\maketitle

\begin{abstract}
We compute the bulk motions of cosmic voids, using a $\Lambda$CDM
numerical simulation considering the mean velocities of the dark matter 
inside the void itself and that of the haloes in the surrounding
shell.
We find coincident values of these two measures in the range \mbox{$\sim$
300-400\kms}, not far from the expected mean peculiar velocities of
groups and galaxy clusters.
When analysing the distribution of the pairwise
relative velocities of voids, we find a remarkable bimodal behaviour
consistent with an excess of both systematically approaching and
receding voids.
We determine that the origin of this bimodality resides in the void
large scale environment, since once voids are classified into
void-in-void (R--type) or void-in-cloud (S--type), R--types are found
mutually receding away, while S--types approach each other.   
The magnitude of these systematic relative velocities account for more
than 100\kms, reaching large coherence lengths of up to 200\hmpc.
We have used samples of voids from the Sloan Digital Sky Survey Data
Release 7 (SDSS-DR7) and the peculiar velocity field inferred from
linear theory, finding fully consistent results with the simulation
predictions.
Thus, their relative motion suggests a scenario of a sparkling
Universe, with approaching and receding voids according to their local
environment.
%
\end{abstract}

\begin{keywords}
   Cosmology: large scale structure of Universe -- Cosmology:
   observations
\end{keywords}

\section{Introduction}
\label{S_intro}

The distribution of galaxies at large scales reveals a complex
structure where cosmic voids, underdense nearly spherical regions
devoid of galaxies, arise as matter flows away from primordial
underdense perturbations toward filaments and walls. 
New insights on the distribution and dynamics of these structures to
unprecedented large cosmological scales, will be available by upcoming
surveys such as HETDEX \citep{hill_hetdex_2008} or Euclid
\citep{euclid_2013}.
Since it is widely believed that voids are weakly clustered and have
negligible velocities, their study is promising for the field of
observational cosmology.

Although the large scale galaxy and mass distributions are dominated
by filaments and walls, a larger fraction of the volume of the
Universe is occupied by voids that can be used in several cosmological
tests \citep{lavaux_voids_2010, bos_darkness_2012}, and galaxy
evolution models\citep{Ceccarelli_2008,ricciardelli_star_2014}.
As the Universe evolves and galaxies flow away from voids, the
supercluster-void network emerges \citep{frisch_evolution_1995,
einasto_supercluster_1997, einasto_multimodality_2012}.
The global flows of mass associated with this clustering process are
expected to be significant up to the scales of the largest structures,
vanishing to a random component at larger scales.  
Cosmic bulk flows have been reported in the local
Universe at scales of a few hundred Mpc, 
some of them challenging the $\Lambda$CDM model   
\citep{watkins_consistently_2009,Lavaux_flows_2010,
Feldman_flow_2010,Colin_2011}.
and others in agreement with this structure formation
scenario
\citep{Nusser_flows_2011,branchini_linear_2012,Turnbull_flows_2012,Ma_Pan_2014,Hoffman_cosmic_bulk_2015}.
It is well known that voids are approximately in isotropic expansion,
a direct consequence of their low density contrast in the central
regions \citep{weygaert_bertschinger_1996, padilla_spatial_2005,
ceccarelli_voids_2006}.
However, it has not been studied into detail the bulk velocity of the
void region and that of the surrounding shell of galaxies, the main
aim of this work.


\begin{figure}
\includegraphics[width=0.48\textwidth]{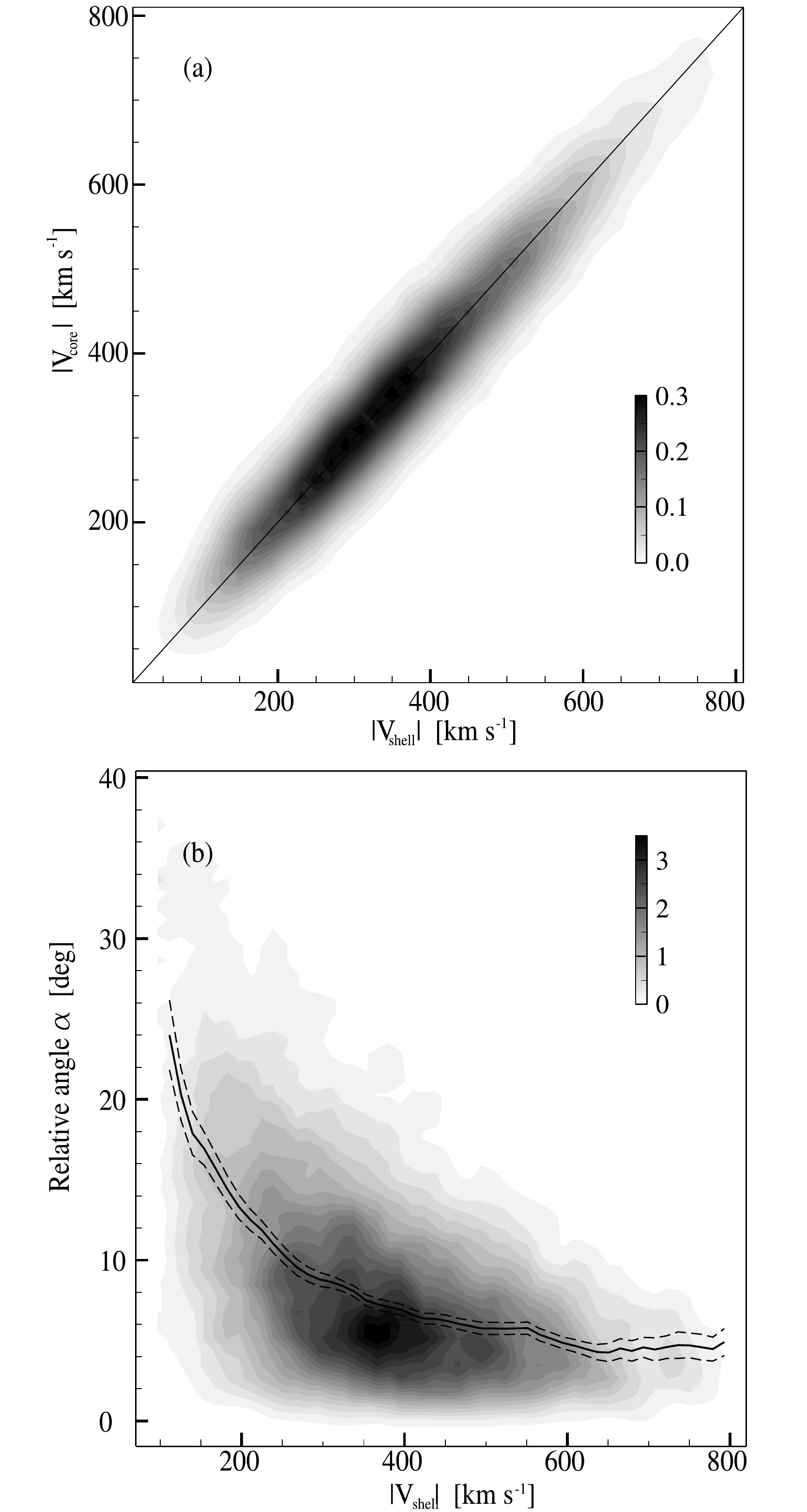}
\caption{
Bulk velocities of void shells and cores in the simulation.
The shell velocity corresponds to the dark matter haloes mean velocity
within $0.8<r/R_{void}<1.2$.
Void core velocity corresponds to the mean of dark matter particles
within $0.8~R_{void}$.
(a) Distribution function of void counts in $|V_{shell}|$,
$|V_{core}|$ bins.
Solid line shows the one-to-one relation, close to the linear fit
results with slope 0.96.
(b) Distribution function of void counts in bins of $|V_{shell}|$ and
the relative angle $\alpha$ between shell and core bulk velocities.
Solid and dashed lines correspond to the median and its standard
error.
\label{F_core}
} 
\end{figure}

\section{Void catalogues}
\label{S_voids}


Cosmic voids can be identified both in simulations and in
observational catalogues.
We use a Voronoi tesselation technique to identify the lowest density
regions, which serve as candidates for void search.
We use a dark matter $N$-body simulation of $1024^3$ particles in a
comoving box of side 1000\hmpc in the standard $\Lambda$CDM scenario.
We select cosmological parameters according to WMAP9 data
\citep{hinshaw_wmpa9_2013}: $\Omega_M=0.279$, $\Omega_\Lambda=0.721$,
$\Omega_b=0.0462$, $h=0.7$, $n=0.972$ and $\sigma_8=0.821$.
The initial conditions were generated using \textsc{MUSIC}
\citep{hahn_music_2011} and the simulation was evolved to $z=0$ with
the public version of \textsc{GADGET-2} \citep{springel_gadget2_2005}.
Dark matter haloes were identified as bound structures using the
\textsc{ROCKSTAR} code \citep{behroozi_rockstar_2013}. The final
catalogue consists of $3983265$ haloes with at least $20$ particles.
To construct the void catalogues we followed the procedure described
in \citet{ruiz_clues_2015}, which is a modified version of the
procedures presented in \citet{padilla_spatial_2005} and
\citet{ceccarelli_voids_2006}.
The density estimation is performed using a Voronoi tessellation on
the halo catalogue, where underdense cells are selected as void center
candidates.
We consider spheres centered in those positions and compute the
integrated density contrast (namely $\Delta(r)$) at increasing values
of radius, then selecting as void candidate the largest sphere
satisfying the condition $\Delta(r) < -0.9$.
After that, the center position is randomly tilted and the sphere is
allowed to grow in order to recenter the void, so that a void of
radius $R_{\rm void}$ is the largest sphere satisfying the underdense
condition and not overlapping with any other underdense sphere.
The final catalogue comprises $13430$ voids in the simulation box with
radii in the range 8--30\hmpc.

In a similar fashion, we identified voids in the observations using
the Main Galaxy Sample of the Sloan Digital Sky Survey Data Release
7 \citep{abazajian_seventh_2009}.
The spectroscopic catalogue comprises in this release 929555
galaxies with a limiting magnitude of \mbox{$r~\leq~17.77$}. 
We adopted volume complete samples with a limiting redshift
$z=0.12$ and a maximum absolute magnitude in the $r-$band of
$M_r=-20.3$.
The limiting redshift of the sample is determined by the dilution
of the sample of galaxies required to achieve statistically
significant results.  
Moreover, the minimum number of
particles for haloes used to identify voids in the
simulation was chosen so that the number density of tracers
in both cases are comparable.  More details on the choice
of parameters that affect the properties of the void sample
are given in \citep{paz_clues_2013}.

\subsection{Linear velocity field in the SDSS}

We have adopted the peculiar velocity field derived from linear theory
by \citet{wang_reconstructing_2012}.  The method is described in
detail in \citet{wang_reconstructing_2009}.
Briefly, these authors use groups of galaxies as tracers of dark
matter halos and its cross correlation function with mass, in order to
estimate the matter density field over the survey domain.
The linear theory relation between mass overdensity and peculiar
velocity is used to reconstruct the 3D velocity field.
To test the accuracy of this method they use a mock catalogue 
that reproduces the main features of the SDSS-DR7 survey. The authors 
compare the actual three--dimensional velocities 
from the simulation box and the reconstructed velocity field in the mock catalogue. They
show that reconstructed velocities are linearly correlated with actual 
velocities. By analysing this correlation as a function of the mock catalogue boundaries,
they infer that the bias in the reconstruction is small in the inner region 
of the survey.
The volume for which this field is reliably determined on this method
is approximately two thirds of the total volume of the SDSS galaxy
sample.  
Thus, we have considered 245 voids, ranging sizes between 5 and
23\hmpc, identified in the observational data restricted to this
smaller region.    
In order to asses the effects of this procedure on the motion of void
shells, we have compared the mean fully non--linear velocities of
haloes within void shells in the simulation and the mean linearized
velocities computed in the same shell volume.
We find that the two vector are typically aligned within 15 degrees
and with moduli differing less than 30\%, showing that linear theory
provides a suitable approximation to the actual void shell velocities.
%


\begin{figure*}
\includegraphics[width=\textwidth]{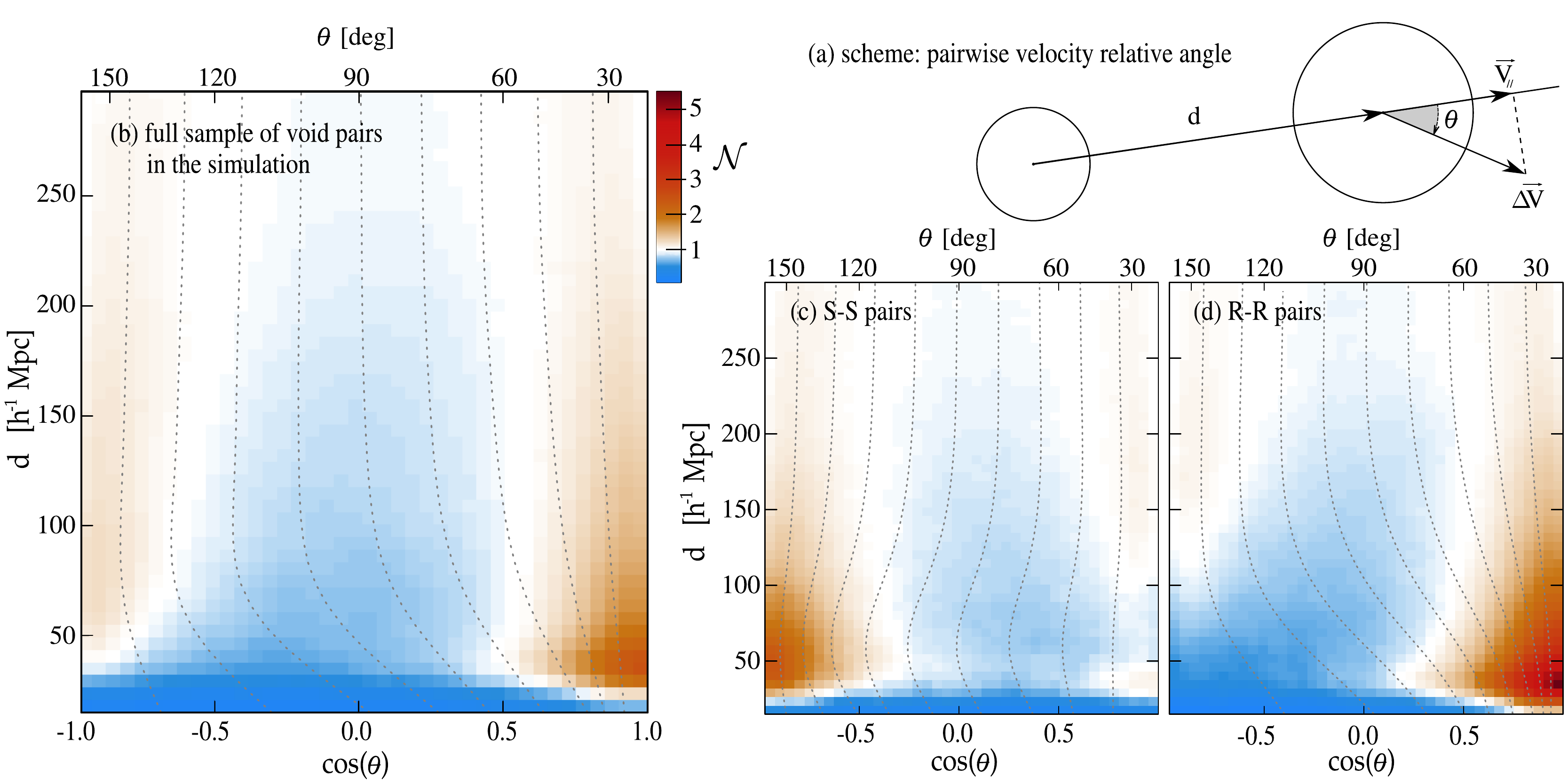}
\caption{
Coherence and bimodality of cosmic void relative motions.
We show $\mathcal{N}$, the number counts of void pairs as a function
of $\cos(\theta)$ and separation, normalized to the expected number
for a uniform distribution of voids without velocity coherence.
The angle $\theta$ is that subtended by the void pairwise velocity
($\vec{\Delta V}$) and the void relative separation ($\vec{\Delta
R}$), as portrayed in panel (a). 
Panel (b) shows the results for all void pairs, panels (c) and (d)
correspond to S-S and R-R pairs, respectively. 
Dotted lines represent the deciles of the distribution of
$\cos(\theta)$.
\label{F_bimodal} 
}
\end{figure*}

\section{Void motions and pairwise velocities}
\label{S_results}

\subsection{Bulk velocities of void shells and cores in the simulation}

In the simulation, we have computed the mean motions of mass
(particles) in the central regions of voids ($r/R_{void}<0.8$) as well
as the mean motions of the haloes in the surrounding shells
($0.8<r/R_{void}<1.2$).
The median values of velocity moduli and relative angle between the
two void velocity measures are shown in Figure \ref{F_core},
indicating that either using particles or haloes, the motion of cosmic
voids is not negligible at all. 
It can be noticed that the dark matter in the void inner region and
the haloes in the surrounding shell, exhibit remarkably similar
velocities, both in magnitude and direction, showing that the void
inner material and the surrounding haloes have a global common motion.
We find that, for 90 per cent of the sample, the direction of the bulk
motion of the inner mass in voids and the haloes in the surrounding
shell agree within 20 degrees, differing in less than $\sim$50\kms in
magnitude.
In consequence, we argue that the use of haloes (or galaxies) in the
shell is suitable to compute the void bulk velocity with the advantage
of the significantly larger number of massive haloes and/or luminous
galaxies than in the void core region.  
We notice that the magnitudes of void peculiar velocities
($\sim$300-400\kms) are comparable to the mean velocity of the highest
density peaks of the large scale structure, such as galaxy clusters.
A more detailed study of the bulk motions of voids will be the subject
of a forthcomming paper.

\subsection{Coherence and bimodality of void relative motions}

In this subsection we analyze whether void bulk motions are randomly
distributed, or their dynamics is related to the large scale structure
and in particular to the void distribution.
With this aim we have analysed the pairwise relative velocities of
voids in the simulation in order to explore the coherence of the
velocity field traced by voids.
We have calculated the number of voids in relative separation,
$d=|\vec{\Delta R}|$, and $\cos(\theta)$ bins, where $\theta$ is the
angle between the void relative velocity and the void relative
separation vectors.
The geometry of this calculation is shown in panel (a) of Figure
\ref{F_bimodal}.
Thus, when the members of a void pair are mutually approaching
(receding), the cosine of the angle between the relative velocity
$\vec{\Delta V}=\vec{V_2}-\vec{V_1}$ and the relative position
$\vec{\Delta R} = \vec{R_2} - \vec{R_1}$ is negative (positive), while
for a non coherent motion it is expected a uniform distribution of
$\cos(\theta)$.
In panel (b) of Figure \ref{F_bimodal} we show the number counts of
void pairs as a function of $\cos(\theta)$ and separation. We
normalize these number counts with the expectation from a uniform
distribution of voids without velocity coherence. It can be clearly
seen the presence of two peaks, particularly within void separations
lesser than 200\hmpc, showing the presence of two populations with
voids mutually receding and approaching.

\begin{figure*}
\includegraphics[width=\textwidth]{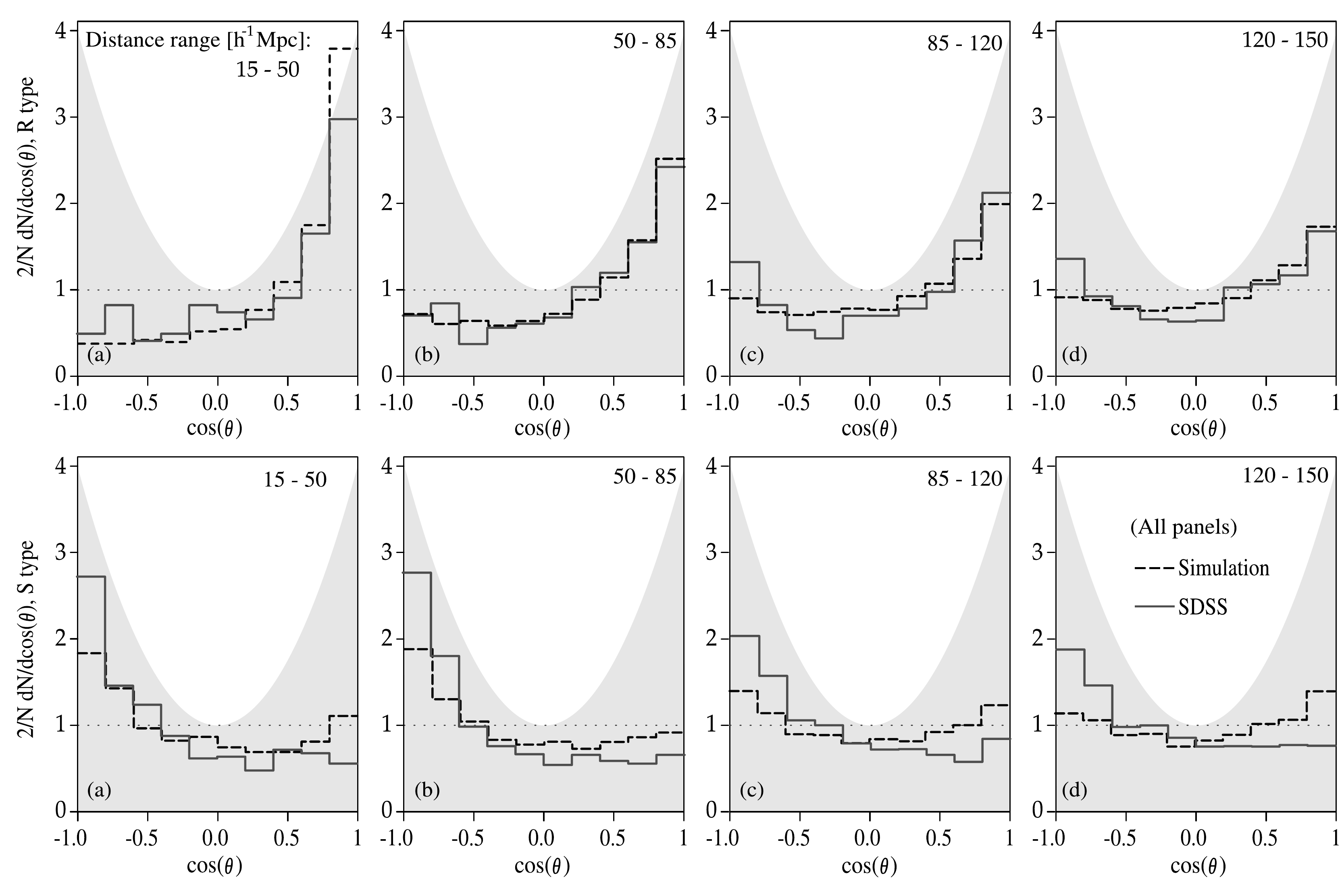}
\caption{
   Bimodality of relative motions in the simulation and
   SDSS data.
   The histograms of $\cos(\theta)$ correspond to void pair
   separations in the ranges 15-50\hmpc (panels a),
   50-80\hmpc (panels b), 80-120\hmpc (panels c) and
   120-150\hmpc (panels d).
   Upper (bottom) panels correspond to
   R-type (S-type) void pairs in the 
   simulation box (dashed lines) and the observations (solid
   lines).
   We show for reference a quadrupolar distribution
   with arbitrary normalization.
   Histograms are normalized to show the excess of void pairs with
   respect to the expectation from a random distribution.
\label{F_cos} 
}
\end{figure*}

We have analysed this somewhat unexpected result within the framework
of the seminal work of \citet{sheth_hierarchy_2004}, who studied the
evolution of cosmic voids in the context of a hierarchical scenario of
structure formation.
They introduced the so called void-in-void and void-in-cloud schemes,
where voids are affected by the surrounding distribution of mass.
This model agrees with the observed dynamics in SDSS voids
\citep{paz_clues_2013}: depending on their environment, some of the
voids shrink while others are in expansion.
Given this strong dichotomy of void dynamics, we have attempted at
linking the void pairwise velocity coherence shown in panel (b) of
Figure \ref{F_bimodal} to their local environment.

Following the methods described in \citet{ceccarelli_clues_2013}, we
have defined voids with continuously rising integrated radial density
profiles (R-type, a proxy for void-in-void), and void samples that
exhibit an overdense shell (S-type, a proxy for void-in-cloud).
By construction, these R- and S-type void samples exhibit different
dynamical properties \citep{paz_clues_2013, ruiz_clues_2015}. 
Here, we further explore the possible connection between void types
and the bimodality of the relative motions shown in panel (b) of
Figure \ref{F_bimodal}.
Remarkably, the observed bimodal behavior can be completely understood
in terms of the R/S void type classification. As it can be seen in
panels (c) and (d), S-type void pairs are systematically approaching
each other, and conversely R-type voids are mutually receding.
We notice that the relative fraction of R/S void types depends on void
radius \citep{ceccarelli_clues_2013}, small voids are typically
S-types while large voids are mainly R-type.  Thus the prominence of
the two peaks seen in this Figure, would vary according to the sample
mix of void sizes.
As expected, at sufficiently large separations, the results are
consistent with uncorrelated motions of void pairs.

In order to analyse this behaviour in the observations, we use void
samples taken from the SDSS data together with the velocity field
derived by \citet{wang_reconstructing_2012}.
Figure \ref{F_cos} shows the distribution of the cosine of the
relative angle between the pairwise void velocity and the relative
separation vector for both observational and simulated voids. As seen
in this Figure, we also obtain a bimodal behavior for observational
voids finding a very good agreement between observations and the
simulation results.  According to the environment classification of
voids, there is a strong relative motion coherence which gently
declines towards large separations.
We also computed the pairwise velocities of haloes and the relative angles
between pairs, finding that the motions of haloes also exhibit a
bimodal behaviour.
This indicates that the coherence pattern is not a peerless feature of
voids, but a ubiquitous consecuence of the large scale distribution of matter.
However, voids offer the oportunity to analyze this coherence pattern in its
double regime of approaching and receeding systems according to their
environment.
We will further discuss these results in an upcoming paper.

\begin{figure}
\includegraphics[width=\columnwidth]{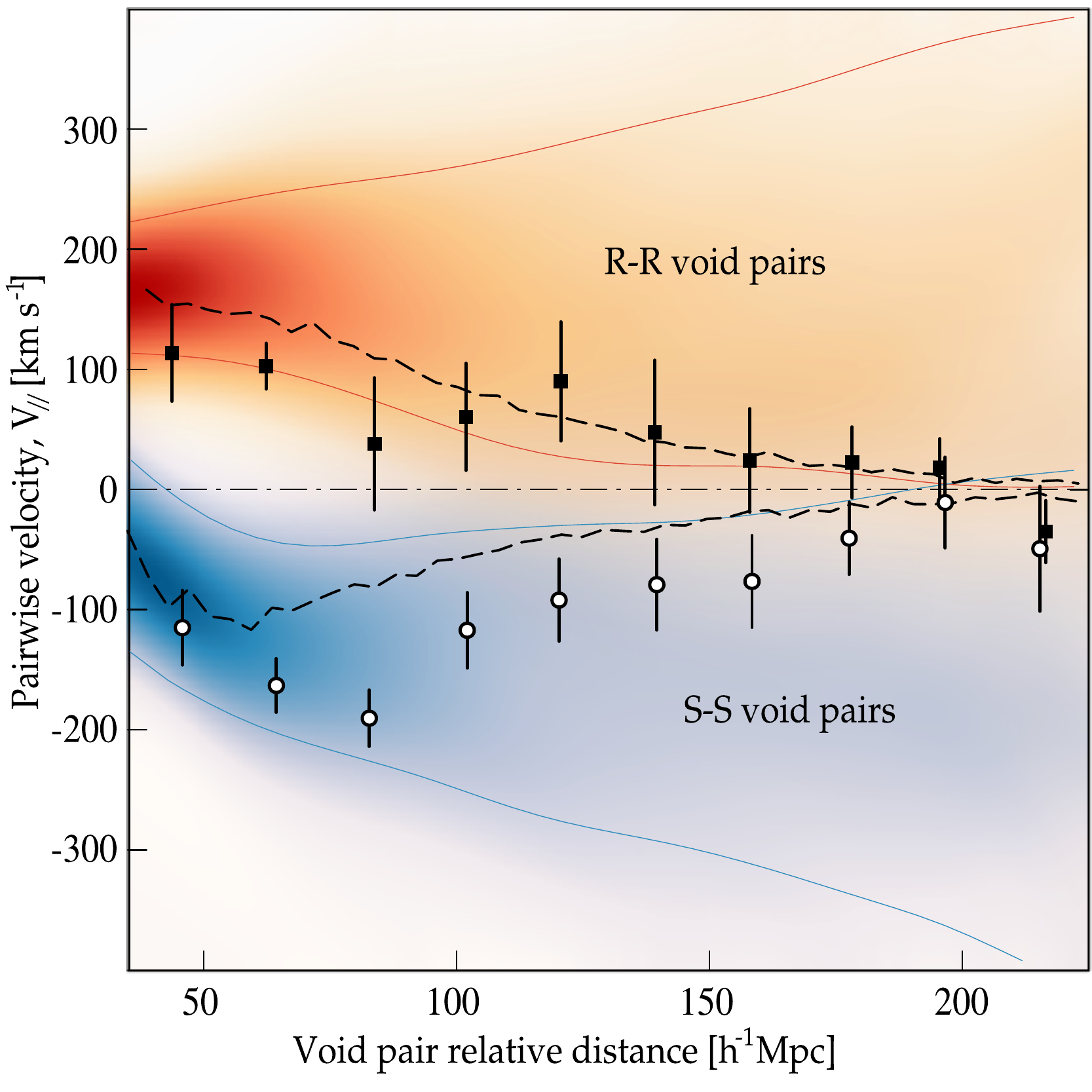}
\caption{ 
   Void pairwise velocity ($V_{\vparallel}$) as a
   function of relative separation in the SDSS data and the
   simulation.
   Points represent SDSS values of $V_{\vparallel}$ and its
   uncertainty computed through Jackknife resampling.
   Dashed lines show the median values of $V_{\vparallel}$ in the
   full-box simulation.
   Red and blue regions represent the cosmic variance obtained from
   subsets of the simulation with similar volume and geometry than the
   SDSS survey for R- and S-types, respectively. 
   The 0.16 and 0.84 quantiles of the distribution of $V_{\vparallel}$
   values are shown in solid lines.
\label{F_vel} 
}
\end{figure}

In order to assess the magnitude of the coherent motions in the
observed Universe, we show in Figure \ref{F_vel} the mean pairwise
velocity values of the SDSS voids as a function of void relative
separation, for R-R void pairs (filled squares) and S-S voids pairs
(empty circles).
The points correspond to the mean values of
$V_{\vparallel}=|\vec{\Delta V}|\cos(\theta)$ and the error bars
represent the uncertainties derived through Jackknife resampling of
the void data.
The colour density map correspond to the results of R-R and S-S void
pairs in 500 different sub--boxes taken at random from the simulation
constrained to have similar SDSS volumes and geometry, thus accounting
for cosmic variance in the SDSS region.
Blue (red) shaded areas correspond to the number density of curves for
S-S (R-R) void pairs.
The thin blue and red lines correspond to the 0.16 and 0.84 quantiles
of the distribution of $V_{\vparallel}$, for S-S and R-R void pairs,
respectively.
The thick dashed lines correspond to the full simulation box results
for R-R and S-S pairs.
As it can be seen in this Figure, the observational results are
entirely consistent with the prediction of the $\Lambda$CDM model.
Voids behave either receding or approaching each other according to
their R/S-type classification with velocities of the order of
\mbox{100--150\kms} up to 200\hmpc separation.

\section{Discussion}
\label{S_discussion}

At a first glance, the coherence and non-negligible motions of voids
have important effects for observational cosmology.
In particular, cosmological tests such as the Alcock-Paczynski
\citep{lavaux_voids_2010,sutter_ap_2012} or the Integrated Sachs-Wolfe
effect \citep{granett_isw_2008, papai_isw_2011,hernandez-monteagudo_signature_2013, ilic_detection_2013,
cai_cmb_2014}, depend on a proper modeling of the velocity and density
fields surrounding voids.
The existence of coherent void motions in large cosmological volumes
can also impact on cosmological tests using voids in forthcoming
large-scale surveys.

Aside the implications on cosmological parameter estimation, the
motion of voids and its coherence, seen here by first time on
observations and the concordance cosmology simulation, is itself
important.
These results could help to understand the formation and evolution of
structure in the larger scales that have been studied so far.
The magnitude of the velocities involved, around 400\kms, and the
coherence of such motions up to scales of 200\hmpc, is expected to
drive the formation of future large scale structures in the Universe
\citep{Boss_2014}.
Another interesting implications of our results can be found in
studies of the bulk flow motion of galaxies and clusters at large
scales
\citep{Nusser_flows_2011,Turnbull_flows_2012,Ma_Pan_2014,Hoffman_cosmic_bulk_2015}.
The measurements of a non-negligible bulk flow at scales around a few
100\hmpc is source of tension between real data and the $\Lambda$CDM
model.
A large amplitude of bulk flow (e.g. $\sim$400\kms) can be taken as an
indication of the presence of significant density fluctuations at very
large scales \citep{watkins_consistently_2009}. 
The existence of such fluctuations could be incompatible with
cosmological models derived from Cosmic Microwave Background probes
\citep{watkins_consistently_2009,Lavaux_flows_2010,Feldman_flow_2010,Colin_2011}.
Our results contribute to the field by adding a possible scenario for
large scale flows based on the concordance cosmological model.
The correlated mutual void velocities observed here can be conducing
at some extent the motion of clusters and galaxies at intermediate
densities, in filaments or walls.
%

Summarizing, this letter reports by first time on the significant
motions of cosmic voids as a whole and study the coherence pattern
associated to the void velocity field up to large cosmological scales,
both in simulations and observations.          
By embracing the idea that voids are moving objects, our results have
an important impact on future cosmological test planed for the next
generation of large volume surveys.                                             
Void coherent bulk velocities, with a bimodal dynamical population of
mutually attracting or receding systems, contribute to imprint large
scale cosmic flows, shaping the formation of future structures in the
Universe.


\section*{Acknowledgments}

This work was partially supported by the Consejo Nacional de
Investigaciones Cient\'{\i}ficas y T\'ecnicas (CONICET), and the
Secretar\'{\i}a de Ciencia y Tecnolog\'{\i}a, Universidad Nacional de
C\'ordoba, Argentina. 
We thank the anonymous referee for useful suggestions that
significantly increase the correctness and quality of this work.
We thank Dr. Mario A. Sgr\'o for kindly providing the numerical
simulation and halo catalogue. 
Plots are made using R software.  This research has made use of NASA's
Astrophysics Data System.

\bibliographystyle{mnras}

\bsp  
\label{lastpage}

\end{document}